\documentclass[twocolumn]{aastex631}

\usepackage{graphicx}   
\usepackage{amsmath}    
\usepackage{mathrsfs}    
\usepackage{bm}
\usepackage{amssymb}    
\usepackage{diagbox}    
\usepackage{subfigure}
\usepackage{enumerate}
\usepackage{float}
\usepackage{ulem}
\usepackage{url}
\shorttitle{sdO/B from AGB CE channel}
\shortauthors{Z. Li et al.}

\begin{document}
\title{A new route to massive hot subdwarfs: common envelope ejection from asymptotic giant branch stars}

\correspondingauthor{Xuefei Chen, Zhenwei Li}
\email{cxf@ynao.ac.cn, lizw@ynao.ac.cn}

\author[0000-0002-1421-4427]{Zhenwei Li}
\affiliation{Yunnan Observatories, Chinese Academy of Sciences, Kunming, 650216, People's Republic of China}
\affiliation{Key Laboratory for the Structure and Evolution of Celestial Objects, Chinese Academy of Science, People's Republic of China}
\affiliation{International Centre of Supernovae, Yunnan Key Laboratory, Kunming, 650216, People's Republic of China}

\author{Yangyang Zhang}
\affiliation{Zhoukou Normal University, East Wenchang Street, Chuanhui District, Zhoukou, 466001, People's Republic of China}


\author{Hailiang Chen}

\affiliation{Yunnan Observatories, Chinese Academy of Sciences, Kunming, 650216, People's Republic of China}
\affiliation{Key Laboratory for the Structure and Evolution of Celestial Objects, Chinese Academy of Science, People's Republic of China}
\affiliation{International Centre of Supernovae, Yunnan Key Laboratory, Kunming, 650216, People's Republic of China}


\author[0000-0002-6398-0195]{Hongwei Ge}

\affiliation{Yunnan Observatories, Chinese Academy of Sciences, Kunming, 650216, People's Republic of China}
\affiliation{Key Laboratory for the Structure and Evolution of Celestial Objects, Chinese Academy of Science, People's Republic of China}
\affiliation{International Centre of Supernovae, Yunnan Key Laboratory, Kunming, 650216, People's Republic of China}


\author{Dengkai Jiang}

\affiliation{Yunnan Observatories, Chinese Academy of Sciences, Kunming, 650216, People's Republic of China}
\affiliation{Key Laboratory for the Structure and Evolution of Celestial Objects, Chinese Academy of Science, People's Republic of China}
\affiliation{International Centre of Supernovae, Yunnan Key Laboratory, Kunming, 650216, People's Republic of China}

\author{Jiangdan Li}

\affiliation{Yunnan Observatories, Chinese Academy of Sciences, Kunming, 650216, People's Republic of China}
\affiliation{Key Laboratory for the Structure and Evolution of Celestial Objects, Chinese Academy of Science, People's Republic of China}
\affiliation{International Centre of Supernovae, Yunnan Key Laboratory, Kunming, 650216, People's Republic of China}

\author[0000-0001-5284-8001]{Xuefei Chen}
\affiliation{Yunnan Observatories, Chinese Academy of Sciences, Kunming, 650216, People's Republic of China}
\affiliation{Key Laboratory for the Structure and Evolution of Celestial Objects, Chinese Academy of Science, People's Republic of China}
\affiliation{International Centre of Supernovae, Yunnan Key Laboratory, Kunming, 650216, People's Republic of China}

\author[0000-0001-9204-7778]{Zhanwen Han}

\affiliation{Yunnan Observatories, Chinese Academy of Sciences, Kunming, 650216, People's Republic of China}
\affiliation{Key Laboratory for the Structure and Evolution of Celestial Objects, Chinese Academy of Science, People's Republic of China}
\affiliation{International Centre of Supernovae, Yunnan Key Laboratory, Kunming, 650216, People's Republic of China}
\affiliation{University of the Chinese Academy of Science, Yuquan Road 19, Shijingshan Block, 100049, Beijing, People's Republic of China}





\begin{abstract}
  The hot subdwarf O/B stars (sdO/Bs) are known as extreme horizontal branch stars, which is of great importance in stellar evolution theory. The sdO/Bs are generally thought to have a helium-burning core and a thin hydrogen envelope $(M_{\rm env }<0.02M_\odot)$. In the canonical binary evolution scenario, sdO/Bs are considered to be the stripped cores of red giants. However, such a scenario cannot explain the recently discovered sdO/B binary, SMSS J1920, where the strong Ca H$\&$K lines in the spectrum are found. It suggests that this binary is likely originated from the recent ejection of common envelope (CE). In this {work}, we proposed a new formation channel of massive sdO/Bs, namely sdO/Bs produced from a CE ejection process with an asymptotic giant branch (AGB) star (hereafter AGB CE channel). We constructed the evolutionary model of sdO/Bs and successfully explained most of the important observed parameters of the sdO/B star in SMSS J1920, including the evolutionary age, sdO/B mass, effective temperature, surface gravity and surface helium abundance.
  The minimum sdO/B mass produced from the AGB CE channel is about $0.48M_\odot$. The evolutionary tracks in $\log T_{\rm eff}-\log g$ plane {may explain a fraction of the observational samples} with high-$\log T_{\rm eff}$ and low-$\log g$. Considering wind mass-loss of sdO/Bs, the model could produce helium-rich hot subdwarfs with $\log (n_{\rm He}/n_{\rm H})\gtrsim-1$.  

\end{abstract}




\section{Introduction}
\label{sec:1}

Hot subluminous stars are one type of extreme horizontal branch star. According to the spectral type, hot subluminous stars are also classified as O/B-type hot subdwarfs (sdO/Bs). In the standard frame, sdO/Bs are identified as helium-burning stars with thin hydrogen envelopes ($<0.02M_\odot$). There are many sdO/B binaries that have been detected with short orbital periods from hours to days (see \citealt{kruckow2021} for a recent catalog of close binaries). Some short orbital period sdO/B binaries are supposed to be candidates of type Ia supernova progenitors (e.g. \citealt{maxted2000,vennes2012,pelisoli2021}). The sdO/B binaries are also one type of gravitational wave (GW) sources for future space-borne GW detectors \citep{kupfer2018,kupfer2023}, such as Laser Interferometer Space Antenna (LISA; \citealt{LISA2023}), TianQin \citep{TianQin}, and TaiJi \citep{TaiJi}. 
Owing to the ubiquity, sdO/B stars play an important role 
{in studying} the Galactic structure and evolution, as well as the ultraviolet excess in elliptical galaxies (\citealt{altmann2004,hanz2007}; see \citealt{heber2009,heber2016} for the review). 

The outstanding features in the sdO/B surface are the chemical peculiarities, in particular for the He abundance. The differences in He abundance provide excellent probes for understanding the formation and evolution of hot subdwarfs (e.g., \citealt{edelmann2003,nemeth2012} {and references therein}). Recently, \citet{luo2021} identified 1587 hot subdwarf stars with atmospheric parameters, and the surface He abundance, $\log (n_{\rm He}/n_{\rm H})$, of hot subdwarfs distributed in a large range ($-4<\log (n_{\rm He}/n_{\rm H})<3$). The atmospheres of most sdBs are 
{helium-poor}, while the sdOs show a variety of helium abundances, ranging from He-poor to extreme He-rich atmospheres. There are several models attempting to explain the anomalies of He abundance. For example, the hot subdwarfs with He-deficient atmospheres are thought to be caused by the diffusion processes \citep{greenstein1967}, and those He-rich hot subdwarfs may be produced from double helium white dwarf mergers \citep{zhangx2012,schwab2018} or single star via internal mixing in the hot flasher scenario \citep{miller2008}. Unfortunately, there is no unified formation theory that can simultaneously explain the properties of He abundance distribution \citep{heber2016}. In other words, the diversity of He abundance may imply a variety of formation channels for hot subdwarfs \citep{leiz2023}. 

It is generally believed that hot subdwarfs are formed through binary interactions \citep{webbink1984}. There are three main binary scenarios for hot subdwarfs: the common envelope (CE) ejection, the stable Roche lobe overflow, and the double helium WD merger. The first two channels lead to hot subdwarfs in binary systems, and the last channel results in single hot subdwarfs \citep{han2002,han2003}. The key point for 
{forming} hot subdwarfs is the He core ignition. For a sdO/B star in the binary system, the He core can be ignited when the donor starts mass transfer close to the tip of RGB (or at earlier stages for massive donors; \citealt{wuy2018}). Most sdO/Bs in the observations can be successfully explained in this scenario (hereafter, RGB channel). However, there are some hot subdwarfs show characteristics obviously in conflict with the RGB channel. For example, hot subdwarfs from the RGB channel generally are He-deficient and thus cannot explain the He-rich hot subdwarfs \footnote{{The late core flash in the RGB model may produce sdO/Bs enriched in helium (e.g. \citealt{miller2008,Naslim2013,Dorsch2019}). However, such scenario has not been verified in the observations (see \citealt{geier2022} for detailed discussions).} }\citep{heber2016}. The situation has been aggravated due to the recent discovery of SMSS J1920 \citep{lij2022}. SMSS J1920 is a Roche-lobe filling sdO/B and white dwarf binary with an orbital period of $3.495\;\rm h$, where the sdO/B star has a mass of $\sim 0.55M_\odot$. The orbital period of SMSS J1920 is so short that it is very likely produced from the CE ejection processes. Moreover, the strong Ca H$\&$K lines with a blueshift of $\sim 200\;\rm km\;s^{-1}$ in the spectrum suggests that the binary likely originated from the recent ejection of CE. However, if the hot subdwarf in SMSS J1920 is produced from the RGB channel, the time from the birth of the sdO/B star to the current state (the time elapsed since the CE ejection) is about several $10^7$ yr. Therefore, the scenario conflicts with the deduction of the recent CE ejection event of SMSS J1920. \citet{lij2022} then proposed a new formation channel of hot subdwarfs, i.e., hot subdwarf produced from a CE ejection process with an asymptotic giant branch (AGB) star (AGB CE channel). Different from the RGB star, an AGB star contains a large CO core, helium/hydrogen burning shells, and a hydrogen envelope. In this {work}, following the idea first suggested in \citet{lij2022}, we investigate the properties of hot subdwarfs formed from the AGB CE channel and explore the implications of the observations. 

The remainder of this {work} is structured as follows. In Section \ref{sec:2}, we introduce the basic inputs and assumptions in our models. The main results, including the comparisons between simulated models and the observations, are presented in Section \ref{sec:3}. The summary and conclusion are given in Section \ref{sec:4}.


\section{Model inputs and methods}
\label{sec:2}

The simulations are performed with the state-of-the-art stellar evolution code Modules for Experiments in Stellar Astrophysics (MESA, version 12115; \citealt{paxton2011,paxton2013,paxton2015,paxton2018,paxton2019}). Two cases of metallicity are investigated: $Z=0.02$ for the field stars (Population I stars) and $Z=0.001$ for the halo stars (Population II stars). We adopted the opacity table OPAL type II and mixing-length parameter of 1.8. To avoid the possible existence of ``breathing pulses'', we employed the ``predictive'' mixing scheme for the subdwarf models, as done in \citet{bauer2021}. The ``predictive'' mixing scheme allows the convective core to grow during He core burning without requiring convective overshoot \citep{paxton2018,ostrowski2021}. In the default model, we do not consider the stellar wind of hot subdwarf for convenience. However, the winds have a negligible effect on the surface He abundance of hot subdwarf stars and will be discussed in details in Section \ref{sec:3.3}. 

In the AGB CE ejection channel, the progenitor of a hot subdwarf fills its Roche lobe at the AGB. The binary may enter the CE phase {(depending on the critical mass ratios as functions of the donors' masses and evolutionary stages; see \citealt{geh2010,geh2015,geh2020,geh2023}).} If the CE is successfully ejected by the released energy of orbital shrinkage, a sdO/B binary is born. The CE process is rather complicated and cannot yet be simulated by MESA. {Very recent works aim to determine the CE eject efficiency accurately \citep{geh2022,geh2023b,2022MNRAS.517.2867H,2022MNRAS.513.3587Z,2023MNRAS.518.3966S}.} For simplicity, the CE process is replaced with a high mass-loss rate wind ($\gtrsim 10^{-3}M_\odot\;\rm yr^{-1}$), as done in many previous works (e.g. \citealt{han2002,xiong2017}). We first evolved a single star to the AGB (where the CE begins) and artificially removed the envelope with a high mass-loss rate. The mass loss is stopped when the hydrogen envelope mass drops below the set value. The initial stellar masses are in the range of $1-5M_\odot$ in steps of $0.1M_\odot$. The 
{remaining} envelope mass is a 
{somewhat} uncertain parameter, so we considered several different values of envelope masses, i.e., from $0.001$ to $0.01M_\odot$. {Our inlist files are available at \dataset[doi:10.5281/zenodo.10547534]{https://doi.org/10.5281/zenodo.10547534}.}

\begin{figure}
    \centering
    \includegraphics[width=\columnwidth]{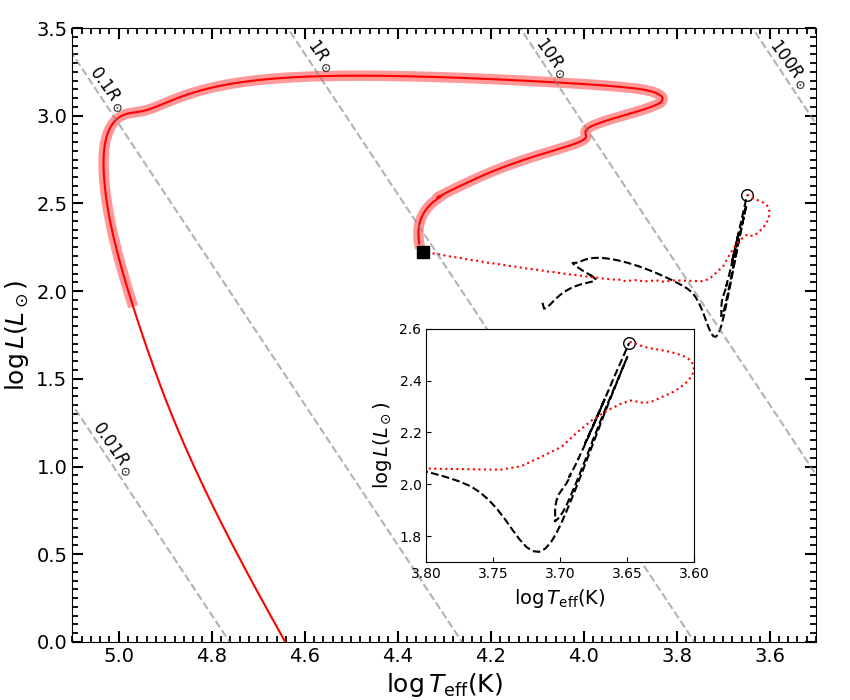}
    \caption{The main steps in constructing the sdO/B model. The black dashed line is for the evolutionary track of a $3.0M_\odot$ star with $Z=0.02$ evolving from MS to the AGB (see the inset for clarity). Then the CE phase happens (black open circle), and the envelope is assumed to be removed with a high mass loss rate until the required envelope mass is reached (here, $0.01M_\odot$ envelope is retained). The track mimicking the CE phase is shown in a red dotted line. The sdO/B is formed with a mass of $0.55M_\odot$ at the black solid square. The thick red line represents the case of $\log {L_{\rm He}}>0$.}
    \label{fig:1}
\end{figure}

In Figure \ref{fig:1}, we present an example to illustrate the details of the simulations. We first evolve a 3.0$M_\odot$ MS star with $Z=0.02$. After the He core burning, it ascends to the AGB. The CE is assumed to happen at the position of the black open circle. Then we removed the envelope with a mass loss rate around $10^{-3}M_\odot\;\rm yr^{-1}$, until the 
{remaining} H envelope\footnote{The boundary of H envelope is defined by hydrogen mass fraction of 0.01.} mass is $0.01M_\odot$. The mass of {the} produced hot subdwarf star is 0.55$M_\odot$, with a $0.33M_\odot$ CO core initially. The sdO/B expands first due to the He shell burning. In the subsequent evolution, the radius decreases, and the luminosity of sdO/B is mainly supported by the shell H-burning. The phase of $\log L_{\rm He}>0$ is shown in the thick line, where $\log L_{\rm He}$ is the He burning luminosity. After the end of He- and H-shells burning, the sdO/B star finally becomes a CO WD. 

\section{Results}
\label{sec:3}

\begin{figure}
    \centering
    \includegraphics[width=\columnwidth]{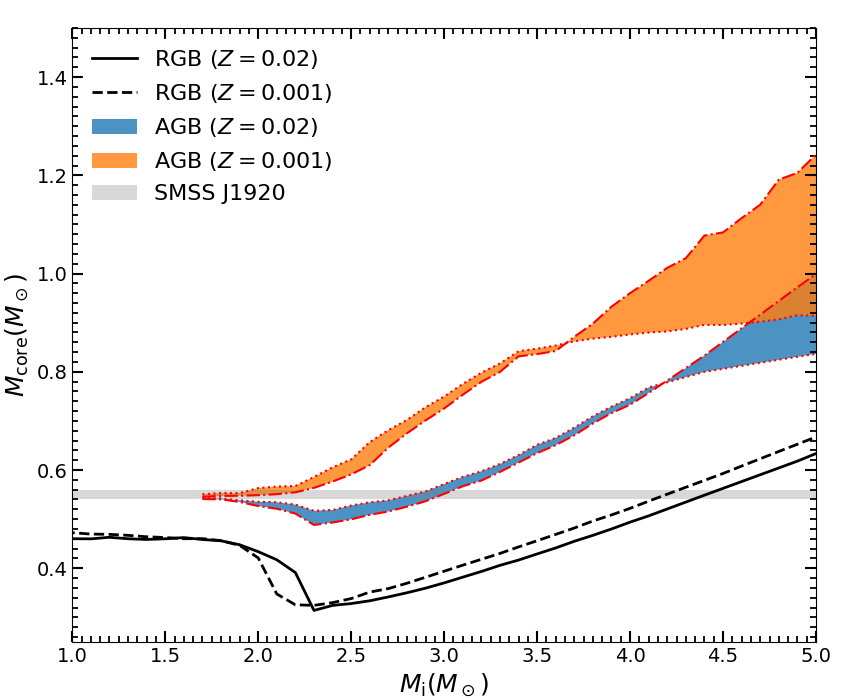}
    \caption{The mass ranges of He core with different stellar masses. The black solid and dashed lines represent the He core mass for a star at the top of RGB with metallicity of $Z=0.02$ and $Z=0.001$, respectively. The blue and yellow hatched regions are for He core masses of sdO/Bs from AGB CE channel with metallicity of $Z=0.02$ and $Z=0.001$, respectively.  
    The red dash-dotted lines represent the He core mass for a star at the AGB, and the radius is just over the maximum radius of the star at the RGB. The red dotted lines are for stars at the TP-AGB. For high-mass stars (massive than $~4.2M_\odot$ for $Z=0.02$ and $~3.5M_\odot$ for $Z=0.001$), the He core mass decreases before the TP-AGB due to the second dredge-up process.} 
    \label{fig:2}
\end{figure}

\subsection{Parameter space of sdO/Bs}
\label{sec:3.1}

To find the possible sdO/B mass in the AGB CE channel, we evolve several single stars with mass in the 
$1 - 5M_\odot$ range. The calculations 
stop as the star enters the thermal pulse-AGB (TP-AGB) phase. If a star fills its Roche lobe and enters into the CE phase at the AGB stage, the radius of the AGB star should be larger than the maximum RGB radius of this star. Otherwise, the mass transfer process happens for a star at the RGB. Therefore, we obtained the helium core mass at the AGB stage when the radius is just above the maximum RGB radius, as shown in red solid lines of the blue ($Z=0.02$) and yellow ($Z=0.001$) hatched regions. The helium core mass at the TP-AGB is shown in red dashed lines. For high-mass stars (massive than $~4.2M_\odot$ for $Z=0.02$ and $~3.5M_\odot$ for $Z=0.001$), the decrease of He core mass before the TP-AGB is caused by the second dredge-up process. The helium core masses for stars at the top RGB are also presented for comparison, as shown in black lines. 

Star with a mass $\lesssim 1.7M_\odot$ possesses a degenerate core at RGB, leading to a large radial expansion before the ignition of He core. Therefore, there is no solution to produce hot subdwarfs in the AGB CE channel for donors with mass $\lesssim 1.7M_\odot$, i.e., the radius of a star at AGB (before the TP-AGB) is always less than the maximum RGB radius. Only stars massive than $1.7M_\odot$ can produce hot subdwarfs through the AGB CE channel. The helium core masses for stars at the AGB are generally larger than {those} of RGB stars. The hot subdwarfs produced from the AGB CE channel are relatively more massive. The minimum helium core mass in the AGB CE channel is $0.48M_\odot$ for $Z=0.02$ and is $0.54M_\odot$ for $Z=0.001$. So, we conclude that the AGB CE channel cannot produce sdO/B stars with masses less than $0.48M_\odot$. \citet{han2003} adopted the binary population synthesis method and found that most sdO/Bs in the RGB channel have masses less than $0.48M_\odot$. It suggests that the mass of sdO/B is an important parameter to distinguish the formation channels. In the recent observations, \citet{leiz2023} obtained the mass distributions for 664 single-lined hot subdwarfs identified in LAMOST and found that there are more sdO/B stars with masses larger than $0.48M_\odot$ when compared with the theoretical mass distribution in \citet{han2003}. {Similarly, \citet{Schaffenroth2022} also found the sdO/B mass distribution sightly extends to higher mass.} Then, we speculate that a part of massive hot subdwarfs originated from the AGB CE channel. Other methods to distinguish the hot subdwarfs from different channels are addressed in Section \ref{sec:3.3}. 

\begin{figure}
    \centering
    \includegraphics[width=\columnwidth]{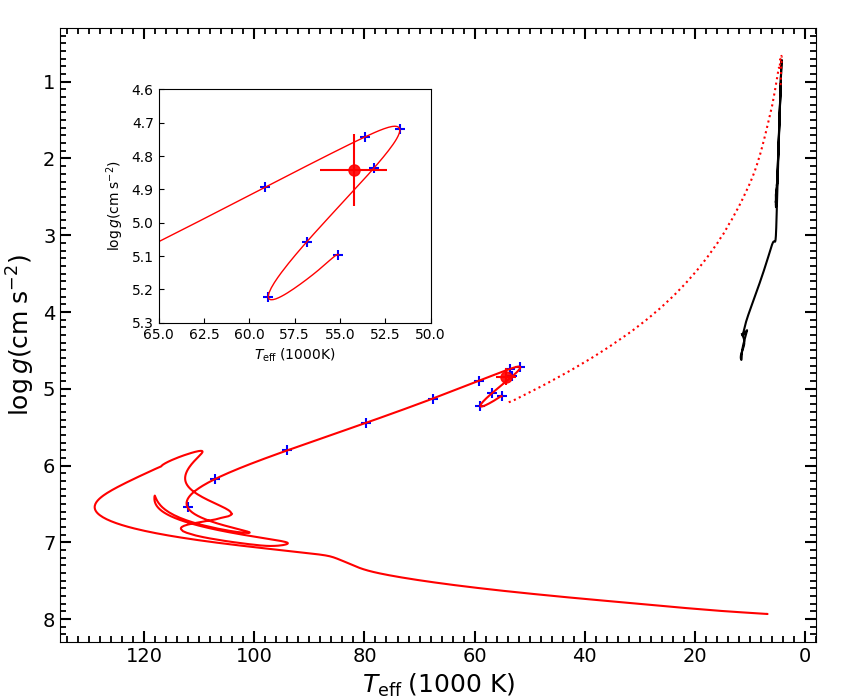}
    \caption{Constructing the evolutionary model of SMSS J1920. The main steps are similar to that introduced in Figure 1. The initial progenitor mass is $1.8M_\odot$ with $Z=0.001$. The mass of the produced sdO/B is $0.553M_\odot$, with an envelope mass of $0.003M_\odot$ and CO core mass of $0.50M_\odot$. The red circles are for SMSS J1920. The time intervals between adjacent pluses are $10^4\;\rm yr$.} 
    \label{fig:3}
\end{figure}

\subsection{Case study of SMSS J1920}
\label{sec:3.2}

SMSS J1920 is the third known Roche lobe-filling hot subdwarf and WD binary found in the observations. The hot subdwarf of SMSS J1920 has a high effective temperature of $\log T_{\rm eff} = 54240\pm1840\;\rm K$ and a relatively low surface gravity of $\log g = 4.841\pm0.108$. The mass of sdO/B star is $0.55\pm0.01M_\odot$, which is {larger} than the canonical mass of hot subdwarf from the RGB channel (typical mass of $0.46M_\odot$, \citealt{han2002}). The outstanding features of strong Ca H$\&$K lines with a blueshift of $\sim 200\;\rm km\;s^{-1}$ in the spectrum suggest that the binary likely originated from the recent ejection of CE (an ejection age of $\sim 10^4\;\rm yr$; see more details in \citealt{lij2022}). 

Based on the observation parameters of SMSS J1920, we try to construct the evolutionary model of the sdO/B star. SMSS J1920 is found in the halo, and then we adopt the metallicity of $Z=0.001$ here. According to the produced helium core mass in the AGB CE channel of Figure \ref{fig:2}, the progenitor mass of sdO/B in SMSS J1920 is in the range of $1.7-2.1M_\odot$. The properties of sdO/Bs are significantly affected by the He core mass and the remaining hydrogen envelope mass. In our best model, we adopt an initial stellar mass of $1.8M_\odot$ and begin to remove the envelope as the star evolves close to the top of AGB and the He core mass increases to $0.55M_\odot$. The process is stopped when the residual envelope mass is around $0.003M_\odot$, as shown in the dotted line of Figure \ref{fig:3}. The evolutionary track of the sdO/B star is shown in a red solid line, where the time intervals between adjacent pluses are $10^4\;\rm yr$. We note that initially most of He material has been burned as C and O, and only about $0.05M_\odot$ He mass remained in the shell. The decreasing He-burning energy leads to the reduction of sdO/B luminosity at the first $10^4\;\rm yr$. The subsequent evolution of sdO/B is supported by the residue H-burning in the shell. The sdO/B track crosses the observational position of SMSS J1920 within $\sim 3\times 10^4 \;\rm yr$, meaning that the age of sdO/B is about several $10^4\;\rm yr$ after the CE ejection event. It supports the view that SMSS J1920 is formed from a recent CE ejection event with an age around several $10^4\;\rm yr$. 

Different progenitor masses in the range of $1.7-2.1M_\odot$ show rather similar characteristics, and all of them can explain the observational parameters of sdO/B in {SMSS J1920}, which are not presented for clarity. One problem that arises from this scenario is that the MS lifetime for a star with mass $\gtrsim1.7M_\odot$ is smaller than $1.2\;\rm Gyr$, while Galaxy halo generally contains old population stars with typical age of $10-12$ Gyr \citep{jofre2011}. Things have improved if we consider the evolutionary timescale of the progenitor of the WD companion star in SMSS J1920. The fact of a lower WD mass than the sdO/B mass suggests that the sdO/B progenitor gains mass from the WD progenitor via the stable Roche lobe overflow \citep{lizw2023}. Assuming an extreme case of fully conservative mass transfer, the WD progenitor should have a mass larger than $\sim 1.1M_\odot$, i.e. transferring $\sim 0.68M_\odot$ to the companion star, corresponding a MS timescale of $\sim 4.5\;\rm Gyr$. Then, the evolutionary age of SMSS J1920 from two ZAMS stars is supposed to be less than $\sim 5.7\;\rm Gyr$, which is still younger than most of the halo stars. However, some observations suggest that many halo stars may arise from the accretion of satellite dwarf galaxies where those stars are generally younger (e.g. \citealt{helmi2008,helmi2018} and reference therein). Our analysis of the origin of SMSS J1920 may support this view. 

\subsection{Comparison with observations}
\label{sec:3.3}

\subsubsection{Evolutionary tracks in $\log T_{\rm eff}-\log g$ plane}
\label{subsec:3.3.1}

\begin{figure}
    \centering
    \includegraphics[width=\columnwidth]{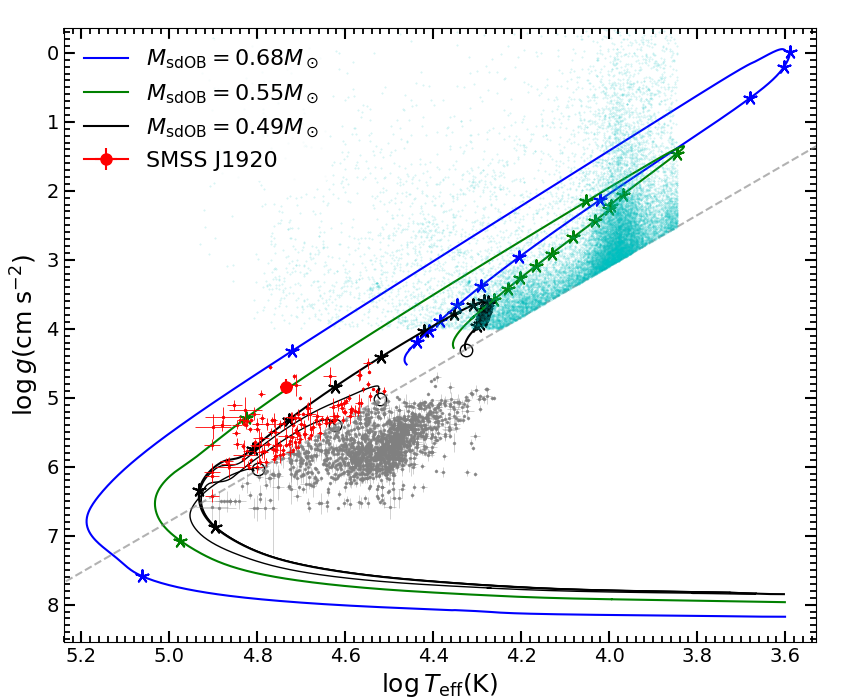}
    \includegraphics[width=\columnwidth]{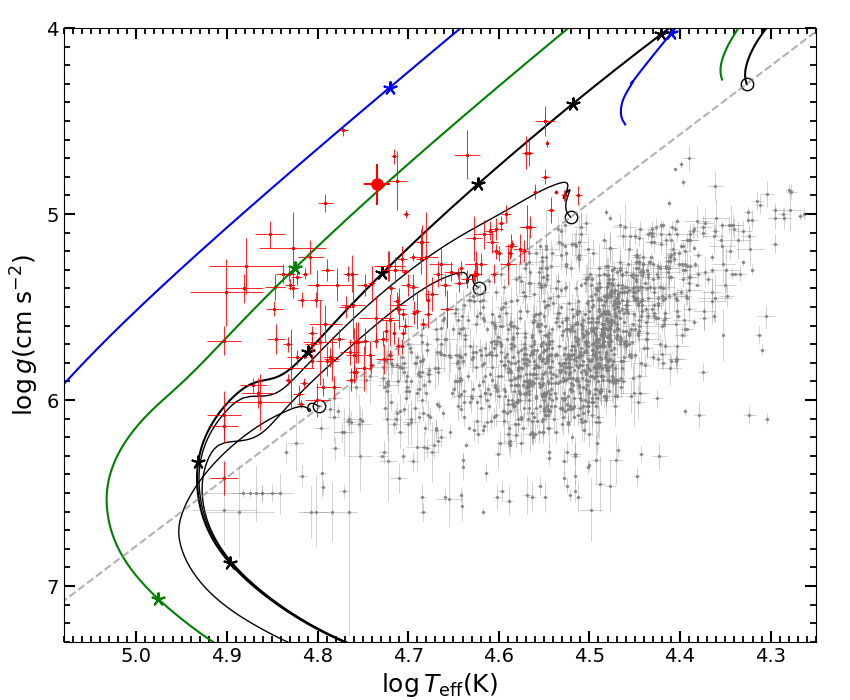}
    \caption{The comparison between the evolutionary tracks of sdO/Bs and the observations in the $\log T_{\rm eff}-\log g$ plane. The whole tracks are shown in the upper panel, and the details compared to the observational sample are shown in the lower panel. The observational hot subdwarf samples are taken from \citet{luo2021}, as shown in red and grey dots with error bars. The cyan dots represent the blue horizontal branch stars taken from \citet{xiang2022}. The typical tracks of sdO/B masses with $0.49,0.55,0.68M_\odot$ and $0.01M_\odot$ envelope mass are presented in thick black, green, and blue lines, respectively. The thin black lines are for $0.48M_\odot$ sdO/Bs with envelope masses of $0.001,0.003,0.005M_\odot$ from the bottom to top, respectively. The grey dashed line across the four starting points of the black lines artificially gives a boundary, above which all of the observation points can be covered by the evolutionary tracks of sdO/Bs. The samples above the grey dashed line are shown in red dots with error bars, and those below the grey dashed line are shown in grey {dots} with error bars. The time intervals between adjacent asterisks are $2\times 10^5\;\rm yr$. } 
    \label{fig:4}
\end{figure}

There are abundant hot subdwarf star samples in the observations, and many of them have precisely detected atmospheric parameters, which can be used to constrain the theoretical model. Recently, \citet{luo2021} identified 1587 hot subdwarf stars with spectra in LAMOST DR7. In Figure \ref{fig:4}, we present the observation sample in the $\log T_{\rm efff}-\log g$ plane, and three simulated evolutionary tracks of sdO/Bs with masses of $0.49,0.55,0.68M_\odot$ are shown in solid thick lines. The envelope masses for all of these three lines are equal to $0.01M_\odot$. It should be noted that the helium core mass of the black thick line is $0.48M_\odot$, which is the minimum core mass obtained in the AGB CE channel (See figure \ref{fig:3}). Therefore, the tracks of sdO/Bs produced in this channel definitely cannot cover all of the observations. 

\begin{figure*}
    \centering
    \includegraphics[width=0.8\textwidth]{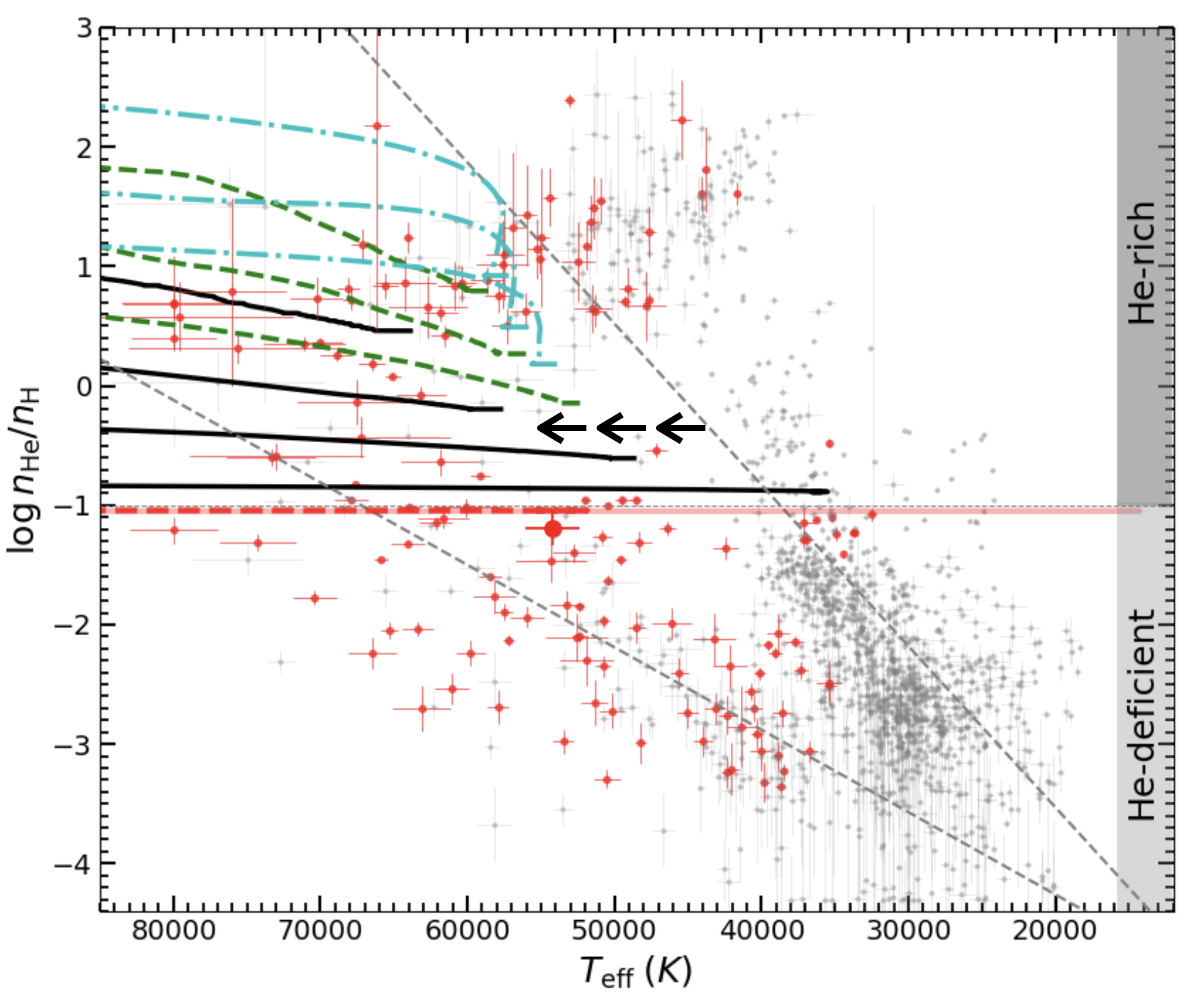}
    \caption{Helium abundance vs. effective temperature for the observations and sdO/B tracks. The red thick lines are for sdO/Bs with metallicity of $Z=0.001$, where the He core mass is $0.55M_\odot$ (solid line for H envelope of $0.005M_\odot$ and dashed line for H envelope of $0.003M_\odot$). Other thick lines are for metallicity of $Z=0.02$. The black solid lines from bottom to top are for sdO/Bs with He core mass of $0.54M_\odot$ and H envelope mass of $0.005,0.003,0.002,0.001M_\odot$, respectively. The green dashed lines are for sdO/Bs with He core mass of $0.62M_\odot$ and H envelope mass of $0.003,0.002,0.001M_\odot$, respectively. The cyan dash-dot lines are for sdO/Bs with He core mass of $0.71M_\odot$ and H envelope mass of $0.003,0.002,0.001M_\odot$, respectively. It is noted that the evolutionary tracks evolves from the right to left, as shown in the black arrows. The observations are taken from \citet{luo2021}, except the SMSS J1920, which is shown in red circles. The red dots are for sdO/B samples that can be covered with the evolutionary tracks in $\log T_{\rm eff}-\log g$ diagram of Figure \ref{fig:4}. The two grey dashed lines are the fitting curves taken from \citet{edelmann2003} and \citet{nemeth2012}. It is clear that our model is able to explain the properties of He-rich hot subdwarfs. The He-deficient atmosphere of sdO/Bs may arise from atomic diffusion and turbulence. See text for more details. } 
    \label{fig:5}
\end{figure*}

To approximately find the boundary of the observational sample that can be explained in our models, we plot several tracks of sdO/Bs with $0.48M_\odot$ helium core and envelope masses of $0.001,0.003,0.005M_\odot$, as shown in thin black lines. The grey dashed line across the four starting points of the black lines artificially gives a boundary, above which all of the observation points can be covered by the evolutionary tracks of sdO/Bs. The timescale during the sdO/B phase (defined as the track above the grey dashed line) is significantly affected by the remaining envelope and sdO/B mass. For the three cases in Figure \ref{fig:4}, the timescales are $4.4\times 10^6, 3.0\times10^6,2.4\times10^6\;\rm yr$ for $M_{\rm sdO/B}=0.49,0.55,0.68M_\odot$, respectively. While the typical lifetime for a canonical sdO/B ($\sim 0.46M_\odot$) is about several $10^8\;\rm yr$ \citep{wuy2018}, about 100 times larger than that of sdO/Bs produced from the AGB CE channel. This may be one of the potential reasons why the number of massive sdO/Bs is relatively rare compared with the canonical sdO/Bs produced from RGB channel \citep{han2003,leiz2023}. We also note the sdO/B stars can expand to several $10R_\odot$ during the {initial} He-shell burning phase, as shown in the upper panel of figure \ref{fig:4}, which may be classified as blue horizontal branch (BHB) stars or giant stars \citep{xiang2022,ju2023}.

\subsubsection{Surface helium abundance}
\label{subsec:3.3.2}

The surface helium abundance is an important quality to trace the evolution of hot subdwarfs. Figure \ref{fig:5} presents the observation sample of \citet{luo2021} in $T_{\rm eff}-\log (n_{\rm He}/n_{\rm H})$ diagram. The two thick-dashed lines are the fitting curves taken from \citet{edelmann2003} for EHB stars (top line) and \citet{nemeth2012} for post-EHB stars (bottom line). The thin dashed line denotes $\log (n_{\rm He}/n_{\rm H}) = -1$, above which is defined as He-rich stars. For a sun-like stars, the surface helium abundance has $\log (n_{\rm He}/n_{\rm H}) = -1$. For stars at the AGB stage, the hydrogen burning happens in the shell, and the resulting sdO/Bs generally have $\log (n_{\rm He}/n_{\rm H})>-1$. The red dots are for the observational sample that can be covered by the evolutionary tracks in the $\log T_{\rm eff}-\log g$ diagram of Figure \ref{fig:4}. During the hot subdwarf evolution, The wind exerts an important effect on the surface abundance. In the following results, we adopt the fitted formula of wind mass-loss rate for hot subdwarfs from \citep{krticka2016}, where the predicted mass-loss rates agree with the values derived from observations. 

We construct several sdO/B models with different He core masses and envelope masses, as shown in Figure \ref{fig:5}. The red thick lines are for sdO/Bs with masses of $0.55M_\odot$ and metallicity of $Z=0.001$, similar to that of SMSS J1920. It is of interest to see that the prediction of He abundance in the model is within the observed error range of SMSS J1920. The wind mass-loss is largely suppressed in the low-metallicity environment \citep{krticka2016}, so the surface He abundance almost has no change, as shown in the red lines. We also simulate several models with metallicity of $Z=0.02$ since most hot subdwarfs are found in the thin disk \citep{luo2021}. The black solid lines from bottom to top are for sdO/Bs with He core mass of 0.54$M_\odot$ and H envelope mass of 0.005, 0.003, 0.002, 0.001$M_\odot$, respectively. The lost mass due to the wind for them is about $0.0005-0.0007M_\odot$. The winds almost have no effect on the surface abundance for sdO/B with a relatively thick envelope, e.g. the case of envelope mass with $0.005M_\odot$. With the decrease of envelope mass, the role of wind becomes non-negligible, especially for sdO/Bs with large masses. The green dashed line and cyan dash-dot line are for sdO/Bs with He core mass of $0.62M_\odot$ and $0.71M_\odot$, respectively. The envelope masses are $0.003,0.002,0.001M_\odot$ from bottom to top for both cases. The high luminosity of massive sdO/Bs leads to a large wind mass-loss rate \citep{krticka2016}. For sdO/Bs with $0.62M_\odot$ He cores (green dashed lines), about $0.002M_\odot$ material are lost due to the wind. It means that for sdO/Bs with initial envelope mass of $0.002M_\odot$ and $0.001M_\odot$, almost all of the envelope material is striped. It is similar to the cases of sdO/Bs with $0.71M_\odot$ He cores (cyan dash-dot lines), where about $0.003M_\odot$ material is lost through the winds. The approximately naked helium core then leads to an extremely high helium abundance in the sdO/Bs surface ($\log (n_{\rm He}/n_{\rm H})\gtrsim 1$). 

The He abundances of our sdO/B models are generally high due to the partial hydrogen burning in the envelope. Then the sdO/B models can naturally explain the distributions of He-rich hot subdwarfs ($\log (n_{\rm He}/n_{\rm H})\gtrsim -1$) with high $\log T_{\rm eff}$, as shown in Figure \ref{fig:5}. Different from the models of double He WD mergers and internal mixing in the hot flasher scenario, our models are proposed to explain the hot subdwarf binaries. Therefore, it would be easy to distinguish the formation channel of He-rich hot subdwarf if there is a companion star. However, in the observations, most He-rich hot subdwarfs seem to be single stars \citep{pelisoli2020,geier2022}, suggesting that He-rich hot subdwarfs produced from AGB CE channel are rare. The possible reason is that the lifetime of massive hot subdwarfs in our models is about two or three orders of magnitude less than that of the normal hot subdwarfs (He-core burning), as discussed in Section \ref{sec:3.3}. We also note that many hot subdwarfs (refer in particular to red dots) with high effective temperature are He-deficient. It means that the wind mass-loss model only is incapable of explaining the properties of He-deficient hot subdwarfs. For these hot subdwarfs, atomic diffusion and turbulence may cause the surface He to sink and lead to He-deficient atmospheres (e.g., \citealt{michaud2011,huh2011}).

\section{Summary and Conclusion}
\label{sec:4}

In this {work}, we investigated the properties of sdO/B stars produced from the AGB CE channel. In this channel, the sdO/B star has a CO core and He- and H-burning shells, and the structure is significantly different from that of the RGB channel. Then, the physical properties for sdO/Bs in the two channels show large differences. The minimum mass of sdO/B star in the AGB CE channel is $\sim 0.48M_\odot$ for $Z=0.02$, and is $0.54M_\odot$ for $Z=0.001$. The mass is generally larger than the canonical mass of the sdO/B star in the RGB channel. Therefore, the mass of a hot subdwarf is an important parameter to distinguish the formation channel. We construct an evolutionary model of the sdO/B star in SMSS J1920, where the sdO/B age is $\sim 10^4\;\rm yr$, supporting the observational view that SMSS J1920 is formed from a recent CE ejection event. We made a comparison between the evolutionary tracks and the observations in $\log T_{\rm eff}-\log g$ plane and the models may explain hot subdwarfs with high-$\log T_{\rm eff}$ and low-$\log g$. Considering the wind mass-loss rate of sdO/Bs, a part of hot subdwarfs with He-rich atmospheres can be naturally explained by the theoretical model. In particular, the prediction of He abundance in our model is within the observed error range of SMSS J1920. So far we could explain most of the important observed parameters of the sdO/B star in SMSS J1920, including the evolutionary age, sdO/B mass, effective temperature, surface gravity and He abundance. The evolutionary timescale of a sdO/B star from AGB CE channel ranges from $\sim 10^5\;\rm yr$ to $\sim 5\times 10^6\;\rm yr$ (depending on the envleope mass and sdO/B mass), which is about 100 times smaller than that of the typical sdO/B lifetime ($10^7$ to several $10^8$ yr). Thus, it may lead to a relatively lower number of sdO/B stars in the AGB CE channel. 

\section*{Acknowledgements}

{We cordially thank the anonymous referee for the detailed comments that improved this manuscript.} We thank Thomas Kupfer for his helpful suggestions at the sdOB11 Conference hold in Armagh. 
This work is supported by the National Key R$\&$D Program of China (Grant Nos. 2021YFA1600403, 2021YFA1600400), the Yunnan Revitalization Talent Support Program—Science $\&$ Technology Champion Project (NO. 202305AB350003), the Natural Science Foundation of China (Grant no. 11733008, 12103086, 11521303, 11703081, 11422324, 12288102, 12073070 12125303), by the National Ten-thousand Talents Program, by Yunnan Province (No. 2017HC018), the Youth Innovation Promotion Association of the Chinese Academy of Sciences (Grant no. 2018076), the Key Research Program of Frontier Sciences of CAS (No. ZDBS-LY-7005), Yunnan Fundamental Research Projects (grant NOs. 202101AU070276, 202101AV070001) and the International Centre of Supernovae, Yunnan Key Laboratory (No. 202302AN360001). We also acknowledge the science research grant from the China Manned Space Project with No. CMS-CSST-2021-A10 and CMS-CSST-2021-A08. The authors gratefully acknowledge the “PHOENIX Supercomputing Platform” jointly operated by the Binary Population Synthesis Group and the Stellar Astrophysics Group at Yunnan Observatories, Chinese Academy of Sciences. 

\software{MESA (v12115; \citealt{paxton2011,paxton2013,paxton2015,paxton2018,paxton2019})}
\bibliographystyle{aasjournal}
\bibliography{sdo.bib}{}
\end{document}